# Direct observation of Anderson localization of matter-waves in a controlled disorder

Juliette Billy[1], Vincent Josse[1], Zhanchun Zuo[1], Alain Bernard[1], Ben Hambrecht[1], Pierre Lugan[1], David Clément[1], Laurent Sanchez-Palencia[1], Philippe Bouyer[1] & Alain Aspect[1]

[1]Laboratoire Charles Fabry de l'Institut d'Optique, CNRS and Univ. Paris-Sud, Campus Polytechnique, RD 128, F-91127 Palaiseau cedex, France

**In 1958, P.W. Anderson predicted the exponential localization[1] of electronic wave functions in disordered crystals and the resulting absence of diffusion. It has been realized later that Anderson localization (AL) is ubiquitous in wave physics[2] as it originates from the interference between multiple scattering paths, and this has prompted an intense activity. Experimentally, localization has been reported in light waves[3,4,5,6,7], microwaves[8,9], sound waves[10], and electron[11] gases but to our knowledge there is no direct observation of exponential spatial localization of matter-waves (electrons or others). Here, we report the observation of exponential localization of a Bose-Einstein condensate (BEC) released into a one-dimensional waveguide in the presence of a controlled disorder created by laser speckle[12]. We operate in a regime allowing AL: i) weak disorder such that localization results from many quantum reflections of small amplitude; ii) atomic density small enough that interactions are negligible. We image directly the atomic density profiles *vs* time, and find that weak disorder can lead to the stopping of the expansion and to the formation of a stationary exponentially localized wave function, a direct signature of AL. Fitting the exponential wings, we extract the localization length, and compare it to theoretical calculations. Moreover we show that, in our one-dimensional speckle potentials whose noise spectrum has a high spatial frequency cut-off, exponential localization occurs only when the de Broglie wavelengths of the atoms in the expanding BEC are larger than an effective mobility edge corresponding to that cut-off. In the opposite case, we find that the density profiles decay algebraically, as predicted in ref 13. The method presented here can be extended to localization of atomic quantum gases in higher dimensions, and with controlled interactions.**

The transport of quantum particles in non ideal material media (e.g. the conduction of electrons in an imperfect crystal) is strongly affected by scattering from the impurities of the medium. Even for weak disorder, semi-classical theories, such as those based on the Boltzmann equation for matter-waves scattering from the impurities, often fail to describe transport properties[2], and fully quantum approaches are necessary. For instance, the celebrated Anderson localization[1], which predicts metal-insulator transitions, is based on interference between multiple scattering paths, leading to localized wave functions with exponentially decaying profiles. While Anderson's theory applies to non-interacting particles in static (quenched) disordered potentials[1], both thermal phonons and repulsive inter-particle interactions significantly affect AL[14,15]. To our knowledge, no direct observation of exponentially localized wave functions in space has been reported in condensed matter.

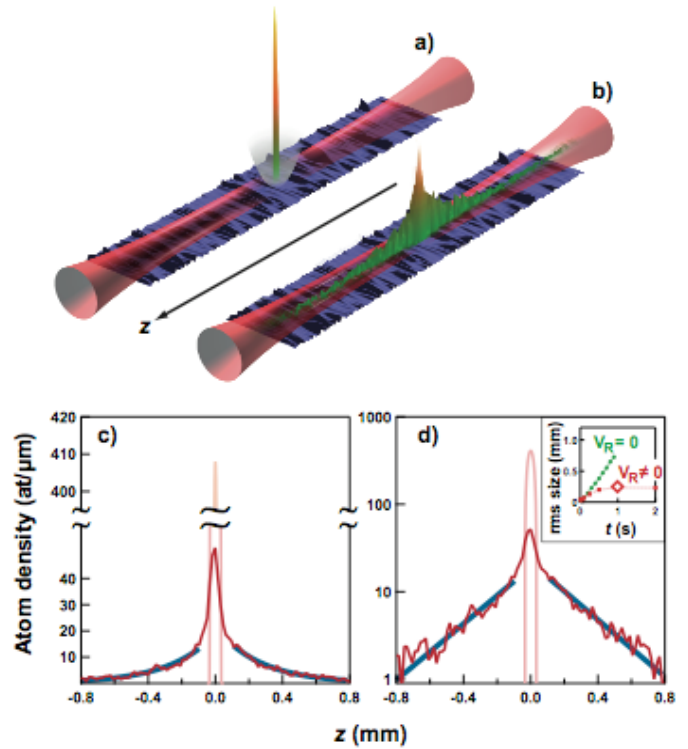

**Figure 1. Observation of exponential localization. a)** A small BEC (1.7 x 10^4 atoms) is formed in a hybrid trap, which is the combination of a horizontal optical waveguide ensuring a strong transverse confinement, and a loose magnetic longitudinal trap. A weak disordered optical potential, transversely invariant over the atomic cloud, is superimposed (disorder amplitude $V_R$ small compared to the chemical potential $\mu_{in}$ of the atoms in the initial BEC). **b)** When the longitudinal trap is switched off, the BEC starts expanding and then localises, as observed by direct imaging of the fluorescence of the atoms irradiated by a resonant probe. On a and b, false colour images and sketched profiles are for illustration purpose, they are not exactly on scale. **c-d)** Density profile of the localised BEC, 1s after release, in linear or semi-logarithmic coordinates. The inset of Fig d (rms width of the profile vs time t, with or without disordered potential) shows that the stationary regime is reached after 0.5 s. The diamond at t=1s corresponds to the data shown in Fig c-d. Blue solid lines in Fig c are exponential fits to the wings, corresponding to the straight lines of Fig d. The narrow profile at the centre represents the trapped condensate before release (t=0).

Degenerate atomic quantum gases can be used to study experimentally a number of basic models of condensed-matter theory, with unprecedented control and measurement possibilities (see ref 16, 17 and references therein). To investigate the behaviour of matter-waves in disordered potentials[18], key advantages of atomic quantum gases are i) the possibility to implement systems in any dimensions, ii) the control of the inter-atomic interactions, either by density control or by Feshbach resonances, iii) the possibility to design perfectly controlled and phonon-free disordered potentials, and iv) the opportunity to measure in-situ atomic density profiles via direct imaging. The quest for evidence of AL of BECs in optical disordered potentials has thus attracted considerable attention in the past years[19,20,21,22]. Experiments using ultracold atoms have shown evidence of dynamical localization associated with a kicked rotor[23,24], which can be considered as a mapping onto momentum-space of the Anderson localization phenomenon. Suppression of one-dimensional transport of BECs has been observed[19,20], but this occured in a regime of strong disorder and strong interactions where localization is due to classical reflections from large peaks of the disordered potential. Here, we report direct observation in real space of one-dimensional localization of a BEC in the regime of AL, i.e. with weak disorder and negligible inter-atomic interactions.

Our experiment (sketched in Fig.1a-b), starts with a small elongated BEC (1.7 x $10^4$ atoms of $^{87}$Rb, which, for the trapping frequencies indicated below, correspond to transverse and longitudinal radii of 3 μm and 35 μm respectively, and a chemical potential of $\mu_{in}/h = 219$ Hz, where $h$ is the Planck constant). The BEC is produced in an anisotropic opto-magnetic hybrid trap. A far off detuned laser beam (wavelength 1.06 μm, to be compared to the resonant wavelength of Rb, 0.78 μm) creates an optical waveguide along the horizontal $z$-axis[25], with a transverse harmonic confinement of frequency $\omega_\perp/2\pi = 70$ Hz. A shallow magnetic trap confines the BEC in the longitudinal direction ($\omega_z/2\pi = 5.4$ Hz). It is suddenly switched off at $t = 0$, and the BEC starts expanding along $z$ in the waveguide, under the effect of the initial repulsive interaction energy. A weakly expelling magnetic field compensates the residual longitudinal trapping of the optical waveguide, so that the atoms can freely expand along $z$ over several millimeters. The expanding BEC can be imaged at any chosen time $t$ after release by suddenly switching off the optical guide and irradiating the atoms with a resonant probe of duration 50 μs. An ultra sensitive EMCCD camera allows us to make an image of the fluorescing atoms with a resolution of 15 μm and a 1D atomic density sensitivity close to 1 atom / μm.

A disordered potential is applied onto the expanding BEC by the use of an optical speckle field produced by passing a laser beam (wavelength 0.514 μm) through a diffusing plate[22]. The detuning from the atomic frequency is so large, and the intensity small enough, that spontaneous photon scattering on the atoms is negligible during the expansion, and we have a pure conservative disordered potential, which extends over 4 mm along $z$. The 3D autocorrelation of the disordered potential, i.e. of the light intensity, is determined by diffraction from the diffusive plate onto the atom location[22]. Transversely, the width of the correlation function (ellipse with axis half-length of 97 μm and 10 μm) is much broader than the size of the atomic matter-wave and we can therefore consider the disorder as one-dimensional for the BEC expanding along $z$ in the waveguide. Along $z$, the correlation function of the disordered potential is $V_R^2 [\sin(\Delta z/\sigma_R)/(\Delta z/\sigma_R)]^2$, where the correlation length, $\sigma_R = 0.26 \pm 0.03$ (s.e.m.) μm, is calculated knowing the numerical aperture of the optics. The corresponding speckle grain size is $\pi \sigma_R = 0.82$ μm. The power spectrum of this speckle potential is non-zero only for $k$-vectors smaller than a cut-off equal to $2/\sigma_R$. The amplitude $V_R$ of the disorder is directly proportional to the laser intensity[22]. The calibration factor is calculated knowing the geometry of the optical system and the constants of $^{87}$Rb atom.

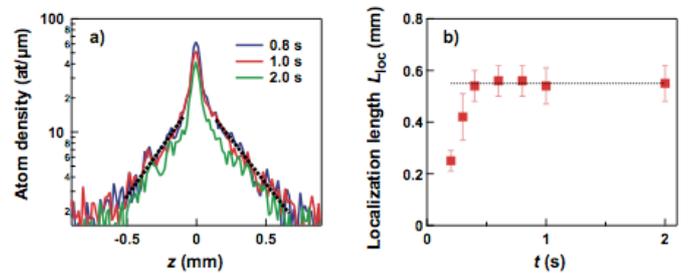

**Figure 2. Stationarity of the localized profile. a**) Three successive density profiles, from which the localization length $L_{loc}$ is extracted by fitting an exponential $\exp(-2|z|/L_{loc})$ to the atomic density in the wings. **b)** Localization length $L_{loc}$ vs expansion time $t$. The error bars indicate 95% confidence intervals on the fitted values (corresponding to ± 2 s.e.m.).

When we switch off the longitudinal trapping in the presence of weak disorder, the BEC starts expanding, but the expansion rapidly stops, in stark contrast with the free expansion case (see inset of Fig.1d showing the evolution of the *rms* width of the observed profiles). A plot of the density profile, in linear and semi logarithmic coordinates (Fig. 1c-d), then shows clear exponential wings, a signature of Anderson localization. We operate here in a regime allowing AL, definitely different from the previous experiments[19,20]. Firstly, the disorder is weak enough ($V_R/\mu_{in} = 0.12$) that the initial interaction energy per atom is rapidly converted into a kinetic energy of the order of $\mu_{in}$ for atoms in the wings, a value much larger than the amplitude of the disordered potential so that there is no possibility of a classical reflection on a potential barrier. Secondly, the atomic density in the wings is small enough (two orders of magnitude less than in the initial BEC) that the interaction energy is negligible compared to the atom kinetic energy. Lastly, we fulfil the criterion stressed in ref 13 that the atomic matter-wave $k$-vector distribution must be bounded, with a maximum value $k_{max}$ smaller than half the cut-off in the power spectrum of the speckle disordered potential used here, i.e. $k_{max} \sigma_R < 1$. The value of $k_{max}$ is measured directly by observing the free expansion of the BEC in the waveguide in the absence of disorder (see Methods). For the runs corresponding to Figs. 1, 2, and 3, we have $k_{max} \sigma_R = 0.65 \pm 0.09$ (± 2 s.e.m.).

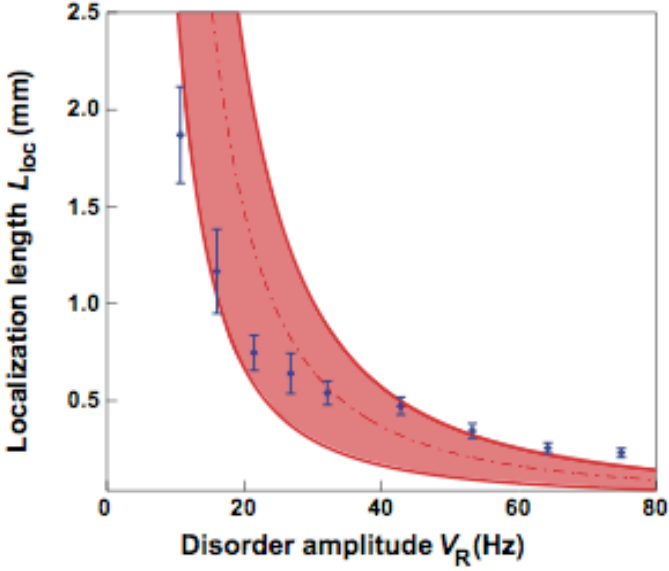

**Figure 3. Localization length *vs* amplitude of the disordered potential.** $L_{loc}$ is obtained by an exponential fit to the wings of the stationary localized density profiles, as shown in Fig. 2. The error bars correspond to a confidence level of 95% of the fit (corresponding to ± 2 s.e.m.). The number of atoms is $N_{at}$ = 1.7 x $10^4$ ($\mu_{in}$ = 219 Hz). The dash-dotted line represents formula (1), where $k_{max}$ is determined from the observed free expansion of the condensate (see Methods). The shaded area represents uncertainty associated with the evaluation of $k_{max}$ and the evaluation of $\sigma_R$. Note that the limited extension of the disordered potential (4 mm), allows us to measure values of $L_{loc}$ up to about 2 mm.

An exponential fit to the wings of the density profiles yields the localization length $L_{loc}$, which we can compare to the theoretical value[13]

$$L_{loc} = \frac{2\hbar^4 k_{max}^2}{\pi m^2 V_R^2 \sigma_R (1 - k_{max}\sigma_R)}, \quad (1)$$

valid only for $k_{max}\sigma_R < 1$ (*m* is the atomic mass). To ensure that the comparison is meaningful, we first check that we have reached a stationary situation where the fitted value of $L_{loc}$ no longer evolves, as shown in Fig. 2. In Fig. 3, we plot the variation of $L_{loc}$ with the amplitude of the disorder, $V_R$, for the same number of atoms, *i.e.* the same $k_{max}$. The dash-dotted line is a plot of Equation (1) for the values of $k_{max}$ and $\sigma_R$ determined as explained above. It shows quite a good agreement between our measurements and the theoretical predictions : with no adjustable parameters we get the right magnitude and general shape. The shaded area reflects the variations of the dash-dotted line when we take into account the uncertainties on $\sigma_R$ and $k_{max}$. The uncertainty in the calibration of $V_R$ does not appear in Fig.3. We estimate it to be not larger than 30 %, which does not affect the agreement between theory and experiment.

An intriguing result of ref 13 is the prediction of density profiles with algebraic wings when $k_{max}\sigma_R > 1$, i.e. when the initial interaction energy is large enough that a fraction of the atoms have a *k*-vector larger than $1/\sigma_R$, which plays the role of an effective mobility edge. We have investigated that regime by repeating the experiment with a BEC containing a larger number of atoms (1.7 x $10^5$ atoms and $\mu_{in}/h$ = 519 Hz) for $V_R/\mu_{in}$ = 0.15. Figure 4a shows the observed density profile in such a situation ($k_{max}\sigma_R$ = 1.16 ± 0.14 (±2 s.e.m.)), and a log-log plot suggests a power law decrease in the wings, with an exponent of 1.95 ± 0.10 (±2 s.e.m.), in agreement with the theoretical prediction of wings decreasing as $1/z^2$. The semi-log plot in inset confirms that an exponential would not work as well. To allow comparison, we present in Figure 4b a log-log plot and a semi-log plot for the case $k_{max}\sigma_R$ = 0.65 with the same $V_R/\mu_{in}$ = 0.15, where we conclude in favour of exponential rather than algebraic tails. These data support the existence of a cross-over from exponential to algebraic regime in our speckle potential.

Direct imaging of atomic quantum gases in controlled optical disordered potentials is a promising technique to investigate a variety of open questions on disordered quantum systems. Firstly, as in other problems of condensed matter simulated with ultra-cold atoms, direct imaging of atomic matter-waves offers unprecedented possibilities to measure important properties, such as localization lengths. Secondly, our experiment can be extended to quantum gases with controlled interactions where localization of quasi-particles[26,27], Bose glass[14,15,28] and Lifshits glass[29] are expected, as well as to Fermi gases and to Bose-Fermi mixtures where rich phase diagrams have been predicted[30]. The reasonable quantitative agreement between our measurements and the theory of 1D Anderson localization in a speckle potential demonstrates the high degree of control in our set-up. We thus anticipate that it can be used as a quantum simulator for investigating Anderson localization in higher dimensions[31,32], first to look for the mobility edge of the Anderson transition, and then to measure important features at the Anderson transition that are not known theoretically, such as critical exponents. It will also become possible to investigate the effect of controlled interactions on Anderson localization.

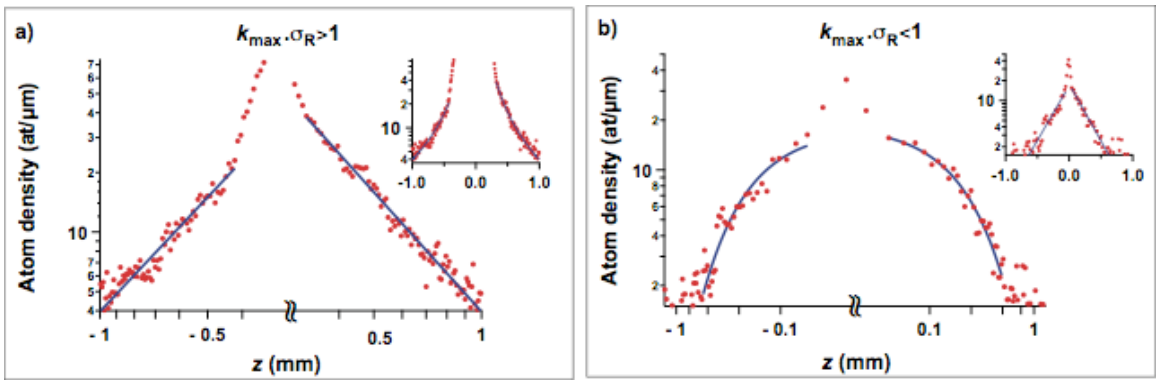

**Figure 4. Algebraic *vs* exponential regimes in a 1D speckle potential.** Log-log and semi-log plots of the stationary atom density profiles showing the difference between the algebraic ($k_{max}\,\sigma_R>1$) and the exponential ($k_{max}\,\sigma_R<1$) regimes. **a)** Density profile for $V_R/\mu_{in}$ = 0.15 and $k_{max}\,\sigma_R$ = 1.16±0.14 (±2 s.e.m.). The momentum distribution of the released BEC has components beyond the effective mobility edge $1/\sigma_R$. The fit to the wings with a power law decay $1/|z|^\beta$ yields β=1.92 +/- 0.06 (±2 s.e.m.) for the left wing and β=2.01 +/- 0.03 (±2 s.e.m.) for the right wing. The inset shows the same data in semi-log plot, and confirms the non-exponential decay. **b)** For comparison, similar set of plots (log-log and semi-log) in the exponential regime with the same $V_R/\mu_{in}$ = 0.15 and $k_{max}\,\sigma_R$ = 0.65 ±0.09 (±2 s.e.m.).

## Methods

**Momentum distribution of the expanding BEC.** In order to compare measured localization lengths to Equation (1), we need to know the maximum value $k_{max}$ of the *k*-vector distribution of the atoms at the beginning of the expansion in the disordered potential. We measure $k_{max}$ by releasing a BEC with the same number of atoms in the waveguide without disorder, and observing the density profiles at various times *t*. Density profiles are readily converted into *k*-vector distributions ($k = \hbar^{-1} m\, dz/dt$). The key point to obtain $k_{max}$ is to determine accurately the position $z_{max}$ of the front edge of the profile. For this, we fit the whole profile by an inverted parabola, which is the expected shape for the 1D expansion of a BEC in the fundamental transverse mode of the waveguide. Actually, the BEC has an initial transverse profile slightly enlarged because of interactions between atoms, but its density rapidly decreases during the expansion, and a numerical calculation with our experimental parameters shows that for expansion times larger than t = 0.2 s, an inverted parabola correctly approximates the atomic density profile, and yields good determination of the front edge position. With this procedure, we measure $z_{max}$ every 0.1s from $t = 0$ to $t = 1$s, and find it proportional to $t$ for $t > 0.2$s. We estimate the uncertainty on $k_{max}$ to about 6 % and 9 % for 1.7 x $10^5$ atoms and 1.7 x $10^4$ atoms respectively.

***Acknowledgements :*** *The authors are indebted to Pierre Chavel, Thierry Giamarchi, Maciej Lewenstein and Gora Shlyapnikov for many fruitful discussions, to Patrick Georges and Gérard Roger for assistance with the laser and to Frédéric Moron, André Villing and Gilles Colas for technical assistance on the experimental apparatus. This research was supported by the Centre National de la Recherche Scientifique (CNRS), Délégation Générale de l'Armement (DGA), Ministère de l'Education Nationale, de la Recherche et de la Technologie (MENRT), Agence Nationale de la Recherche (ANR), Institut Francilien de Recherche sur les Atomes Froids (IFRAF) and IXSEA ; by the STREP programme FINAQS of the European Union and by the programme QUDEDIS of the European Science Foundation (ESF). During the completion of this manuscript, we have been made aware of a related work on BEC behaviour in a 1D incommensurate bichromatic lattice (M. Inguscio, private communication).*